\documentclass[aps,prl,reprint,superscriptaddress,showpacs]{revtex4-1}
\usepackage{dcolumn}
\usepackage{import}
\usepackage{graphicx}
\usepackage{xcolor}
\usepackage{amssymb}
\usepackage{geometry}

\newcommand{\lamu}{\lambda_{\text{u}}}
\newcommand {\E}  {\varepsilon}
\newcommand {\om} {\omega}
\newcommand {\Om} {\Omega}

\newcommand {\Nacc} {N_{\text{acc}}}

\newcommand {\Ld}   {L_{\text{d}}}
\newcommand {\Lp}   {L_{\text{p}}}
\newcommand {\Ltot} {L_{\text{tot}}}

\newcommand {\calA} {{\cal A}}

\newcommand {\meanL} {{\langle L \rangle }}
\newcommand {\meanN} {{\langle N \rangle }}

\newcommand {\ach}  {a_{\text{ch}}}
\newcommand {\Omch} {\Om_{\text{ch}}}

\begin{document}

%%%%%%%%%%%%%%%%%%%%%%%%%%%%%%%%%%
\title{Interplay and specific features of radiation 
mechanisms of electrons and positrons in crystalline undulators}

\author{Alexander V. Pavlov}
\email{a.pavlov@physics.spbstu.ru}
\affiliation{Peter the Great St. Petersburg Polytechnic University, 
Polytechnicheskaya 29, 195251 St. Petersburg Russia}
\affiliation{St. Petersburg State Maritime University, Leninsky ave. 101, 
198262 St. Petersburg Russia}

\author{Andrei V. Korol}
\email{korol@mbnexplorer.com}
\affiliation{St. Petersburg State Maritime University, Leninsky ave. 101, 
198262 St. Petersburg Russia}
\affiliation{MBN Research Center, Altenh\"{o}ferallee 3, 60438 Frankfurt am Main Germany}

%\author{Kirill Agapev}
%\affiliation{Peter the Great St.Petersburg Polytechnic University, 
% Polytechnicheskaya 29, 195251 St.Petersburg Russia}
\author{Vadim K. Ivanov}
\affiliation{Peter the Great St. Petersburg Polytechnic University, 
Polytechnicheskaya 29, 195251 St. Petersburg Russia}

\author{Andrey V. Solov'yov}
\affiliation{MBN Research Center, Altenh\"{o}ferallee 3, 60438 Frankfurt am Main Germany}

%%%%%%%%%%%%%%%%
\begin{abstract}
We predict peculiar changes in spectral distributions of radiation
emitted by ultra-relativistic positrons and electrons
in periodically bent crystals with variation of the bending amplitude.
It is shown that the changes, being sensitive to the projectile energy,
manifest themselves differently for negatively and positively charged projectiles.
We demonstrate that the features observed are due to the interplay of
different radiation mechanisms occurring in periodically bent 
crystals.
The observations are important for design,  practical realization and channeling
experiments with periodically bent crystals
as the key element of the novel light sources.
The analysis presented is based on the grounds of accurate numerical simulations 
of the channeling process.

\end{abstract}


%61.85.+p  Channeling phenomena
%41.60.-m  Radiation by moving charges} \and
%41.75.Ht  Relativistic electron and positron beams} \and
%02.70.Uu  Applications of Monte Carlo methods}}
%07.85.Fv X- and γ-ray sources, mirrors, gratings, and detect
\maketitle

%%%%%%%%%%%%%%%%%%%%%%%%%%%
For several decades, propagation of relativistic charged particles 
in oriented crystals remains in focus of challenging research. 
As predicted by Lindhard~\cite{Lindhard_KDan_v34_p1_1965}, the projectiles 
penetrate large distances moving along a crystallographic direction. 
The channeling phenomenon 
has been receiving a significant interest, both with respect to a fundamental 
theory and the experiments.
% , see, e.g., Ref. \cite{ChannelingBook2014} 
% and the references therein. 
%
In straight crystals, the intensity of radiation emitted by channeling particles 
(the channeling radiation, ChR \cite{ChRad:Kumakhov1976}) exceeds by orders of 
magnitude the background bremsstrahlung (BrS) radiation in the amorphous medium.
In addition to ChR, in crystals bent with a uniform curvature $1/R$ 
the radiation acquires features of the synchrotron radiation (SR)
\cite{Taratin_PhysPartNucl_v29_p1063_1998-English,
SolovyovSchaferGreiner_PRE_v53_p1129_1996,Korol_EtAl_NIMB_v424_p26_2018}. 

Another type of radiation appears in a \textit{crystalline undulator} 
(CU) in which a beam of ultra-relativistic electrons or positrons undergoes 
planar channeling in a periodically bent crystal (PBC) \cite{KSG1998,KSG_review_1999}.
The periodic bending of the planes gives rise to the spontaneous undulator-type 
CU radiation (CUR) which can be generated in the range of wavelengths $\lambda$ 
from 0.1 down to $10^{-6}$ \AA.
Its intensity and characteristic frequencies can be varied
by changing the beam energy, the parameters of bending and the type of a crystal.
The CUR peak brilliance 
can be considered at the level of $10^{25}$ photons/s\, mrad$^2$ mm$^2$\, 0.1\%BW 
in the photon energy range $0.1 - 10$ MeV 
\cite{ChannelingBook2014} which cannot be reached in modern undulators based on magnets
\cite{YabashiTanaka_NaturePhotonics_v11_p12_2017}.
Present technologies are nearly sufficient to meet the conditions needed to achieve
the emission stimulation in a CU \cite{ChannelingBook2014}.
The scheme for practical realization of a CU laser was patented 
\cite{Patent}.
This device will be capable of emitting FEL-type radiation 
with $\lambda = 10^{-3} - 10^{-1}$ \AA, i.e. orders 
of magnitude shorter than in the current XFEL facilities 
\cite{Doerr-EtAl_NatureMethods_v15_p33_2018,Seddon-EtAl_RepProgPhys_v80_115901_720_2017,
SwissFEL_ApplScie_v7_720_2017,Emma-EtAl_NaturePhotonics_v4_015006_2010,
McNeilThompson_NaturePhotonics_v4_p814_2010}).
Light source in the hard-X and gamma ray energy range can open new possibilities 
for various experiments and applications 
\cite{Ledingham-EtAl_Science_v300_p1107_2003,Hajima-EtAl_JNuclScieTechnol_v45_p441_2008,
Weon-EtAl_PRL_v100_217403_2008,MBNExplorer_Book}.
This is a very ambitious goal, which imposes further refinement of the
existing apparatus, new experimental approaches and technologies as well as development of
state-of-the-art theoretical and computational methods.
% To this end, one can recall that it took several decades to convert the initial idea
%  of a FEL \cite{Madey1971} into an operating device.
However, the efforts in this direction will certainly make this field of endeavour 
even more fascinating.

%%%%%%%%%%
The feasibility for the CU scheme was predicted and verified theoretically only
very recently \cite{KSG1998,KSG_review_1999,KSG_review2004,Dechan01,
Tabrizi-EtAl_PRL_v98_164801_2007}.
In these papers, as well as in the subsequent publications 
(see Ref. \cite{ChannelingBook2014}
for the latest review), essential conditions and limitations which must be met 
to make possible the observation of the effect were formulated.
These papers triggered a sharp increase in publications on the subject worldwide, 
so that one can state that this topic represents a new, rich and very promising 
field of research.

In recent years, electron channeling and radiation in bent crystals have been 
extensively studied both theoretically and experimentally
\cite{BackeEtAl_JINST_v13_C04022_2018,
Wistisen_EtAl_EPJD_v71_124_2017,
Wistisen_EtAl_PR-AB_v19_071001_2016,Wistisen_etal_PRL_v112_254801_2014,
UggerhojWistisen_NIMB_v355_p35_2015,
WienandsEtAl_PRL_v114_074801_2015,Backe_EtAl_PRL_115_025504_2015,
BackeLauth_NIMB-v355-p24-2015,
Mazzolari-EtAl_PRL_v112_135503_2014,Bagli-EtAl_EPJC_v74_p3114_2014,Kostyuk_PRL2013,
Korol_EtAl_NIMB_v424_p26_2018,
Korol_EtAl_EPJD_v71_174_2017,Korol_EtAl_NIMB_v387_p41_2016,
Sushko_EtAl_NIMB_v355_p39_2015,
Polozkov_EtAl_EPJD_28_268_2014,BentSilicon_2013,Sub_GeV_2013}.
A set of experiments have  been performed aiming at detecting the CUR. 
The most recent ones include experiments at the MAinzer MIcrotron (MAMI)
\cite{BadEms_p58,BackeEtAl_JINST_v13_C04022_2018}, 
CERN \cite{BadEms_p38} and SLAC \cite{Wienands_Talk_2016}. 
So far, these attempts have not been enrirely conclusive most probably due to a 
not sufficient quality of the probed periodically bent (PB) crystalline structures 
\cite{ThuNhi-EtAl_JApplCryst_v50_p561_2017}. 

%%%%%%%%%%%%%%%%%%%%%%%%%%%%%%%%%%%
% In this Letter, we study the channeling and radiation in a diamond crystal, 
% having been motivated by the ongoing experiments at the 
% MAMI \cite{BackeLauth_BadEms_p63,BackeEtAl_JINST_v13_C04022_2018}. 
In this Letter, we predict peculiar changes in the emission spectra with variation of 
the amplitude $a$ of periodic bending.
These modifications are sensitive to the projectile energy $\E$ and
manifest themselves differently for negative and positive projectiles.
We demonstrate that the features observed are due to the interplay of
different radiation mechanisms occurring in periodically bent 
crystals, namely, the ChR, the CUR, the SR and the coherent BrS.
%
% we study the channeling and radiation in a diamond crystal, 
% having been motivated by the ongoing experiments at the 
% MAMI \cite{BackeLauth_BadEms_p63,BackeEtAl_JINST_v13_C04022_2018}. 

Numerical modeling of the channeling phenomena was performed by means of the 
multi-purpose computer package \textsc{MBN Explorer} 
\cite{MBN_Explorer_Paper,MBN_Explorer_Site}.
The channeling module of the package \cite{MBN_ChannelingPaper_2013} 
allows for multiscale all-atom molecular dynamics simulations of 
the ultra-relativistic projectiles propagation and radiation in various enviroments
including the crystalline one. 
The code was benchmarked for various ultra-relativistic projectiles channeling 
\cite{ChannelingBook2014,MBN_ChannelingPaper_2013,BezchastnovKorolSolovyov2014,
Sushko_EtAl_NIMB_v355_p39_2015,Backe_JINST_v13_C02046_2018,
BackeEtAl_JINST_v13_C04022_2018}.
% Detailed description of theoretical frameworks as well as of the simulation procedure
% is given in Refs. \cite{ChannelingBook2014,MBN_ChannelingPaper_2013}.
The atomistic approach implemented in MBN Explorer,
combined with modern numerical algorithms, advanced computational facilities
and computing technologies makes the predictive power of the software 
comparable or maybe even higher than the accuracy achievable experimentally.
Thus, the computational modeling becomes a tool that could substitute 
(or become an alternative to) expensive laboratory experiments, and thus 
reduce the experimental and technological costs.

In line with the experimental conditions for electrons at MAMI
\cite{BackeEtAl_JINST_v13_C04022_2018}
and with those potentially achievable for positrons 
\cite{Backe_EtAl_NuovoCimC_v34_p175_2011}, 
we simulated the channeling of $\E=270-855$ MeV electrons and positrons in 
$L=20$ $\mu$m thick straight and PB diamond(110) crystals. 
The harmonic bending shape $S(z) = a \cos (2\pi z/\lamu)$ 
was assumed with the coordinate $z$ measured along the incident beam direction. 
The bending period $\lamu$ was fixed at the value of 5 $\mu$m, 
the range of bending amplitude includes the values $a = 1.2$, 2.5, 4.0 \AA. 
The values of $\E, L, \lamu$ and $a$ quoted above correspond to those 
used in recent experiments at MAMI \cite{BackeLauth_BadEms_p63}. 

\begin{figure*}
\includegraphics[width=1.0\textwidth]{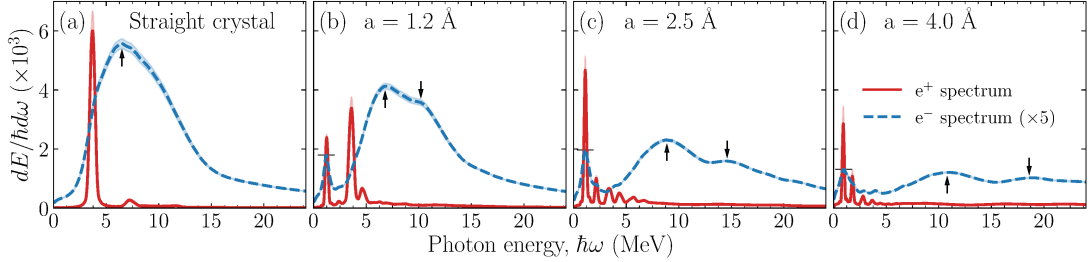}
\caption{
Spectral distributions of radiation by 855 MeV electrons and positrons 
for the straight (a) and PB diamond (110) crystal for 
different bending amplitudes as indicated in (b-d). 
The electron spectra are multiplied by factor 5. 
The upward arrows indicate the maxima of ChR for electrons, the downward 
arrows show the positions of the additinal maxima appearing in the PBCs.
%The spectra correspond to the emission opening angle $\theta_0=0.24$ mrad. 
Grey shading indicates statistical errors due to the finite number of 
simulated trajectories.
For the sake of comparison, we mention that the intensity of the 
background radiation (not shown in the figure) 
due to the incoherent BrS, estimated within the Bethe-Heitler approximation
% \cite{Tsai1974}, 
is 0.025 (in the units $10^{-3}$ used in the figure).
}
\label{Figure01.fig}
\end{figure*}

For each set of the parameters, $N=6000$ trajectories were simulated.
Due to randomness in sampling the incoming projectiles and in positions of the 
lattice atoms due to the thermal fluctuations each simulated trajectory corresponds 
to a unique crystalline environment \cite{MBN_ChannelingPaper_2013}. 
Thus, all simulated trajectories are
statistically independent and can be analyzed to quantify the
channeling process as well as the emitted radiation.
For each trajectory spectral distribution of the emitted radiation 
was calculated within the cone $\theta \leq \theta_0=0.24$ mrad.
% along the beam direction.
% $d E/d(\hbar\om)= N^{-1}\sum_{j=1}^{N}0\int_{0}^{2\pi} d\phi\int_0^{\theta_0}
% \theta d\theta\,(d^3 E/d(\hbar\om) d\Om)$.
% Here $d^3 E/d(\hbar\om)d \Om)$ stands for the spectral-angular distribution
% of radiation from the $j$th trajectory. 
The resulting spectrum was obtained by averaging over \textit{all} trajectories, 
and thus it accounts for contributions of the channeling segments as well 
as of those corresponding the non-channeling regime.
%%%%%%%%%%%%%%%%%%%%%%%%%%%%%%%%%%%%
Figure \ref{Figure01.fig} presents the calculated spectral distributions for 
855 MeV projectiles.
% The data refer to the emission angle  $\theta_0=0.24$ mrad which is much smaller
% than the natural opening angle for the radiation, $\gamma^{-1}\approx 0.6$ mrad
% with $\gamma=\E/mc^2$.
%
% Graph (a) shows the dependences in the straight diamond(110)
% crystal, whereas graphs (b)-(d) correspond to PBC with different bending 
% amplitude as indicated.

%%%%%%%%%%%%%%%%%%%%%%%%%%%%
%%%%%%%%%%%%%%%%%%%%%%%%%%%%%%%

Figure \ref{Figure01.fig}(a) serves reference purposes and illustrates 
well-established features of the emission spectra in 
straight crystals (see, e.g., Ref. \cite{BakEtAl_NPB_v254_p491_1985}). 
For both electrons and positrons the spectra are dominated by peaks of ChR, 
the intensities of which by far exceed that of BrS in the amorphous medium.   
Different character of channeling oscillations by positrons and electrons result in
differences in the peak profiles.
For positrons, nearly perfect harmonic channeling oscillations
give rise to a narrow, undulator-type peak at $\hbar \om \approx 3.6$ MeV. 
% This value correlates with the estimate 
% $2\gamma^2\hbar \Omch = \hbar c\gamma^2(8U_0/\E d^2)^{1/2}$ 
% which one derives within the continuous potential model assuming purely harmonic 
% shape of the interplanar potential ($\Omch$ denotes the frequency of the oscillations).  
% This estimates produces the value of 3.8 MeV if one sets $U_0\approx 20$ eV for the 
% (110) potential well depth in diamond \cite{ChannelingBook2014}.
A much less intensive peak at $\approx 7.2$ MeV corresponds to the 
emission in the second harmonics.
For electrons, strong anharmonicity of the channeling oscillations results in 
significant broadening and lowering of the peak (note that the
electron spectra in are multiplied by factor five).

In the PBCs, Figs. \ref{Figure01.fig} (b)-(d), the spectra exhibit additional 
features some of which evolve differently with the increase in $a$.
% the bending amplitude. 

\noindent
(I) First, there are CUR peaks in the low-energy parts of the spectra.
The most powerful of these correspond to the emission in the
fundamental harmonic of CUR at $\hbar\om_1 \approx 1$ MeV in 
both electron ('-') and positron ('+') spectra.
To be noted is the non-monotonous dependence of the peak intensities, 
$I_{\text{CUR}}^{(\pm)}(a)$, on $a$.
% To be noted is the difference in the peaks width for the two types of projectiles.
% The higher harmonics peaks at $\om_n = n\om_1$ are seen in the positron spectra.

\noindent
(II) The second feature, concerns a strong suppression of ChR in 
the positron spectra as $a$ increases.
Indeed, for periodic bending with $a=1.2$ \AA{} the ChR intensity, 
$I_{\text{ChR}}^{(+)}(a)$, 
drops by a factor of two as compared to the straight crystal, $a=0$, whereas
for larger bending amplitudes the ChR peak virtually disappears.

\noindent
(II) Finally, we mention the evolution of the electron ChR spectrum.
Its intensity $I_{\text{ChR}}^{(-)}(a)$ does not fall off so dramatically 
as for positrons. 
Next, as $a$ increases the ChR peak (marked with the upward arrow in each graph) 
becomes more blue-shifted and there appears additional
structure (the downward arrows) on the right shoulder of the spectrum.

In what follows, the physical explanation of these novel features is provided.

A particle, channeled in a PBC, experiences two types of 
quasi-periodic motions: due to the channeling oscillations and because of 
periodic bending of the crystal planes.
To estimate the dependence of ChR and CUR on the bending amplitude $a$, 
one notices that intensity $I$ of the radiation emitted by a bunch of particles
is proportional to 
(i) average number $\meanN$ of particles participating in the motion,
(ii) average distance $\meanL$ covered by a particle,
(iii) squared Fourier image of the particle's acceleration, which can be written
as $\Om^4 A^2$ with $\Om$ and $A$ standing for the frequency and the 
(average) amplitude of the quasi-periodic motion:
\begin{eqnarray}
I \propto \meanN \meanL \Om^4 A^2\,.
\label{eq.01}
\end{eqnarray}
This general relation can be applied to both electrons and positrons.

Let us first analyze the case of a positron channeling.
In a planar channeling regime, a charged projectile moves along the planar 
direction experiencing a collective action of the electrostatic field of the lattice
atoms \cite{Lindhard_KDan_v34_p1_1965}. 
For a positively charged projectile, the field is repulsive, so that the particle 
is steered into the inter-atomic region and oscillates (channels) in between two 
adjacent crystal planes. 
At some stage, due to the collisions with the crystal constituents,
the transverse energy becomes large enough to allow particle to leave the channeling
mode, i.e. to dechannel. 
The opposite process, the re-channeling, is associated with the capture to the
channeling mode.
In a sufficiently thick crystal, a projectile can experience dechanneling and 
re-channeling several times in the course of propagation.
In this case, the quantity $\meanL$ should account for the length of all channeling
segments in the trajectory. 
However, for relatively thin crystals, $L \ll \Ld$,
(where $\Ld$ denotes the dechanneling length, i.e. the average length of a 
channeling segment in a sufficiently thick crystal,  
e.g., \cite{BiryukovChesnokovKotovBook}), the re-channeling events are rare. 
This feature has been revealed also in a series of recent simulations 
\cite{ChannelingBook2014,Sub_GeV_2013,Korol_EtAl_NIMB_v387_p41_2016,
Korol_EtAl_NIMB_v424_p26_2018,Korol_EtAl_EPJD_v71_174_2017}. 
Using Eq. (1.50) from \cite{BiryukovChesnokovKotovBook} one estimates $\Ld$
for a 855 MeV positron in diamond(110) as ca $500$ $\mu$m
which is much larger than the thickness $L=20$ $\mu$m.
In this limit, the quantity $\meanL$ can be calculated as a mean penetration length
$\Lp$ of the accepted particles, i.e. those captured into the channeling mode at the 
crystal entrance. 
Correspondingly, the quantity $\meanN$ can be associated with
the number of accepted particles, $\Nacc$, which is related to the channel
acceptance $\calA=\Nacc/N$ with $N$ being the total number of particles.
The acceptance is maximum for a straight crystal, and gradually decreases
with increase in the bending curvature $1/R$
due to the action of the centrifugal force \cite{Tsyganov_TM-682_1976}.
In a PBC, the maximum curvature of the cosine bending profile
$1/R_{\max}\approx 4\pi^2a/\lambda_{\rm u}^2$ increases with $a$, leading 
to the decrease in $\Nacc$ as well as in the values of $\Lp$.

%%%%%%%%%%%%%%%%%
\begingroup
\squeezetable
\begin{table}[ht]
\caption{
Acceptance $\calA$, penetration depth of accepted particles
$\Lp$, 
mean squared amplitude of channeling oscillations
$\langle a_{\rm ch}^2\rangle$ (in units of $d^2$, $d=1.26$ \AA{} stands
for the (110) interplanar spacing), 
and total length of channeling segments $\Ltot$ 
for $855$ MeV projectiles in $20$ $\mu$m thick
straight ($a=0$) and PB diamond(110) crystal. 
}
\label{Table_ep-data.C}   
\resizebox{\columnwidth}{!}{ %%%
\begin{tabular}{rrrrrrrrr}
\hline\noalign{\smallskip}
         &\ &  \multicolumn{3}{c}{positrons} & \ &  \multicolumn{3}{c}{electrons}\\ 
$a$ (\AA)&\ & $\calA$&$\Lp$ ($\mu$m)& $\langle a_{\text{ch}}^2\rangle/d^2$ 
         &\ & $\calA$&$\Lp$ ($\mu$m)& $\Ltot$ ($\mu$m) \\ 
\noalign{\smallskip}\hline\noalign{\smallskip}
  0      &\ & 0.96  & $19.3 \pm 0.1$ & $0.060$ &\ & 0.72 & $10.9 \pm 0.3$  & $11.8 \pm 0.3$ \\
1.2      &\ & 0.82  & $19.0 \pm 0.2$ & $0.042$ &\ & 0.48 & $ 7.3 \pm 0.3$  & $ 8.0 \pm 0.2$ \\ 
2.5      &\ & 0.60  & $16.8 \pm 0.3$ & $0.033$ &\ & 0.29 & $ 5.0 \pm 0.3$  & $ 4.4 \pm 0.2$ \\ 
4.0      &\ & 0.24  & $15.1 \pm 0.6$ & $0.010$ &\ & 0.20 & $ 3.5 \pm 0.2$  & $ 2.5 \pm 0.1$ \\ 
\noalign{\smallskip}\hline
\end{tabular}
} %%%
\end{table}
\endgroup
%%%%%%%%%%%%%%%%%

The second and third columns in Table \ref{Table_ep-data.C} summarize 
the data on $\calA(a)$ and $\Lp(a)$ obtained via statistical analysis of the 
simulated trajectories of positrons (the value $a=0$ stands for the straight 
channel). 
These data allows one to qualitatively analyze the dependence of the 
intensity $I_{\text{CUR}}^{(+)}$ of CUR for positrons on $a$.
Using Eq. (\ref{eq.01}) and taking into account the independence on the amplitude
the factor $\Om_{\rm u}^2$ is independent on the amplitude, one writes:
$I_{\text{CUR}}^{(+)}(a) \propto \calA(a) \Lp(a) a^2$.
Here, the product of a decreasing $\calA(a) \Lp(a)$ and an increasing $a^2$ functions
of bending amplitude results in the existence of $a_{\max}$ which ensures the
maximum value of the CUR intensity.

Similar methodology is applicable to analyze the dependence of the ChR intensity
$I_{\text{ChR}}^{(+)}$ on $a$. 
In this case, assuming the harmonic character of the channeling oscillations, 
% i.e. independence of the frequency $\Om_{\rm ch}$ on the amplitude 
% $\ach$ of the oscillations, 
one writes Eq. (\ref{eq.01}) as follows:
$I_{\text{ChR}}^{(+)} \propto \calA(a) \Lp(a) \langle a_{\rm ch}^2\rangle(a)$.
Here, $\langle a_{\rm ch}^2\rangle(a)$ stands for the mean square amplitude
of the channeling oscillations.
This quantity is a decreasing function of $a$. %the bending amplitude.
Indeed, 
% for a straight channel, $a=0$, the maximum amplitude $\ach$ equals 
% approximately to the half width of the channel, $d/2$.
as $a$ increases, the centrifugal force, especially in the vicinity of the points 
of the maximum curvature, drives the projectiles
oscillating with large amplitudes away from the channel resulting in a strong
quenching of channeling oscillations.
The fourth column in Table \ref{Table_ep-data.C} provides the data on 
$\langle a_{\rm ch}^2\rangle$. %in units of $d^2$.
Thus, gradual decrease in all three factors, $\calA$, $\Lp$
and $\langle a_{\rm ch}^2\rangle$ results in a considerable drop in the intensity
of ChR: the value of $I_{\text{ChR}}^{(+)}$ at $a=4$ \AA{} is by a factor 
of 30 less than for the straight channel, which is in accordance with the 
trend seen in Fig. \ref{Figure01.fig}.

%%%%%%%%%%%%%%%%%%%%%%%%%%%%
\begin{figure*}
\includegraphics[width=0.8\textwidth]{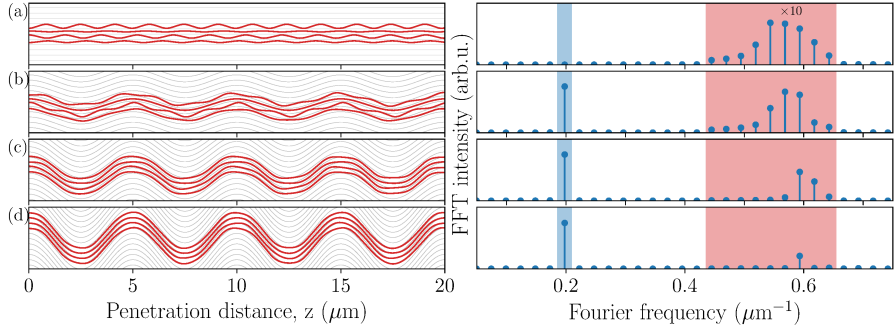}
\caption{
The 2D projections of several {\it exemplary} trajectories of the 
855 MeV positrons (left) channeled through the whole 
$L=20$ $\mu$m thick diamond(110).
Thin lines mark the crystalline channels for positrons. 
Graphs (a)-(d) refer to $a=0$ (the straight crystal),
1.2, 2.5 and 4.0 \AA{}, correspondingly.
The corresponding FFT spectra calculated from {\it all} channeled 
trajectories are presented on the right panel.
The blue (left) vertical strip marks the Fourier signal due to the periodic bending,
the red (right) strip shows the part of the spectrum due to 
channeling oscillation (note the scaling factor 10).
}%
\label{Figure02.fig}
\end{figure*}

To explain the transformations in the emission spectra, let us propose an 
alternative approach based on a quantitative analysis of oscillatory modes of 
the simulated trajectories by means of the fast Fourier
transform (FFT).
This approach allows one to visualize the impact of $a$
on the amplitude of channeling oscillations.

Illustrative left panel in Fig. \ref{Figure02.fig} shows several trajectories 
of positrons which propagate through the whole crystal 
% 20 $\mu$m thick diamond(110) crystal 
in the channeling regime. 
The right panel presents the FFT spectral distributions 
calculated from all channeling trajectories.
In graphs (b)-(d), corresponding to the PBCs, the FFT signals at 
$f=0.2$ $\mu$m$^{-1}$ 
correspond to the motion along the cosine centerline with the period $\lamu=5$ $\mu$m.
These peaks are normalized via division by $a$.

By comparing the spectra within the interval $f=0.45-0.65$ $\mu$m$^{-1}$
(indicated in the red shading) for the straight and PBCs
one sees the modification of the distribution of channeling oscillations 
with respect to the amplitude $\ach \propto\text{FFT}$  
and frequency $\Om_{\rm ch}\propto f$.
In particular, the FFT spectra clearly indicate that with increase in the
bending amplitude the ChR intensity decreases significantly. 

%%%%%%%%%%%%%%%%%%%
\begin{figure*}
\includegraphics[width=0.8\textwidth]{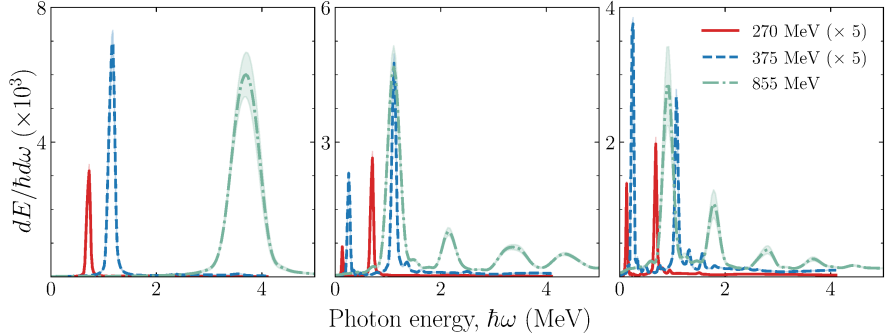}
\caption{
Emission spectra produced by 
$\E=270$, 375 and 855 MeV positrons in straight (left graph) and PB 
(middle and right graphs with $a = 2.5$ and 4.0 \AA{}, correspondingly)
diamond (110) crystals of thickness $L=20$ $\mu$m.
%The spectra correspond to the emission opening angle $\theta_0=0.24$ mrad. 
The shaded areas indicate the statistical errors.
Note the different scales of the vertical axes.
}
\label{Figure03.fig}
\end{figure*}

To conclude the discussion of the positron case, we remark on the dependence of the 
profile of the emitted spectrum on the projectile energy.
Lower values of $\E$ lead to weakening of the centrifugal force $\E/R_{\max}$ which, 
in turn, lessens the suppression of the ChR contribution to the spectrum in a PBC.
Figure \ref{Figure03.fig} illustrates the evolution of the emission spectra 
with $\E$ and $a$. 
Left graph corresponds to the straight crystal and thus shows the 
peaks due to the ChR (note the scaling factor for the $\E=270$ and 375 MeV spectra).
With increase in $a$, middle and right graphs, 
the ChR peak for the 855 MeV projectiles virtually disappears, whereas
for lower $\E$ it is still well pronounced.

Let us now turn to the emission spectra due to electrons, 
see Fig. \ref{Figure01.fig}.
In contrast to the positron case, the dechanneling length $\Ld$ of the 855 MeV electrons 
in straight diamond(110) is less than the crystal thickness $L=20$ $\mu$m.
Our analysis of the trajectories simulated in a much thicker crystal 
(140 $\mu$m) produces the value $\Ld=13.1 \pm 0.2$ $\mu$m \cite{Note1}. %
% \footnote{This result
% is somewhat lower that the experimental data $24.6 \pm 7.9$ $\mu$m 
% of Backe \textit{et al.} \cite{BackeEtAl_JINST_v13_C04022_2018}.
% This difference, however, does not affect the arguments presented in our Letter 
% concerning the evolution of the emission spectra by electrons.}
%%
In a PBC, $\Ld$ is even less being a decreasing function of $a$. 
To estimate the decrease rate, one can consider the values the mean penetration length
$\Lp$ of the accepted electrons given in Table \ref{Table_ep-data.C}. 
As in the positron case, the accepted particles provide the main contribution 
to the low-energy part of the spectrum dominated by CUR.
For higher energies of the emitted radiation, i.e. in the domain of ChR,
the contribution of the re-channeled particles to the electron emission spectrum 
increases.
As shown further in the Letter, the dechanneling--re-channeling dynamics together
with the strong anharmonicity of the channeling oscillations 
lead to modifications in the shapes of the electron ChR spectrum different from those
discussed for the positrons.

To reveal the evolution of the CUR peak in the electron spectra, 
one starts with Eq. (\ref{eq.01}), then follows the arguments outlined above 
when discussing the positron case, and writes 
$I_{\text{CUR}}^{(-)}(a) \propto \calA(a) \Lp(a) a^2$.
This relation is adequate when $\Lp(a)$ contains at least one CU period. 
Using the data from Table \ref{Table_ep-data.C} for $a=1.2$ and $2.5$ \AA{}, one 
explains the differences in the CUR intensities in Figs.  \ref{Figure01.fig}(b) and (d). 
For $\Lp < \lamu$, as it occurs for PB with $a=4$ \AA{}, the radiation becomes
more synchrotron-like which manifests itself in broadening of the peak accompanied
by additional reduction in its intensity.
These features are seen in Fig.  \ref{Figure01.fig}(d).

%%%%%%%%%%%%%%%%%%%
\begin{figure}
\includegraphics[width=0.235\textwidth]{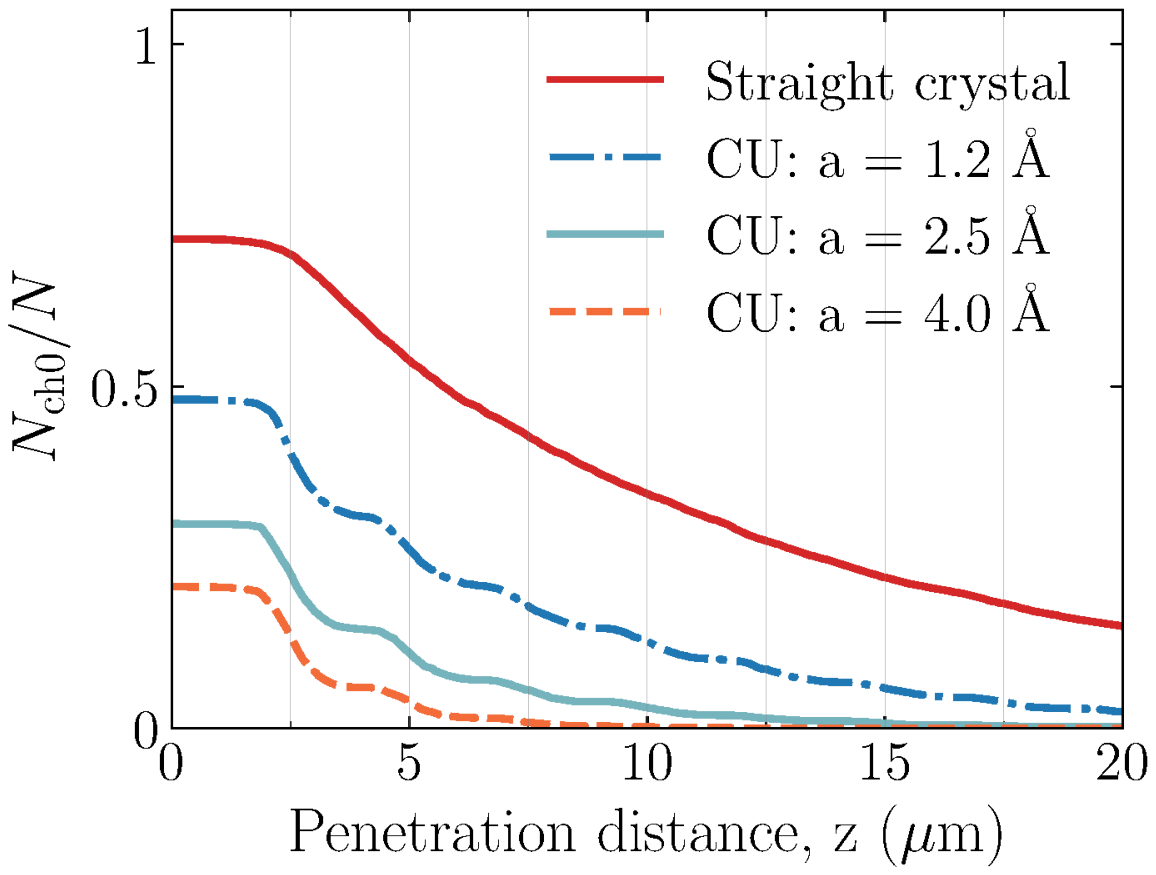}
\includegraphics[width=0.235\textwidth]{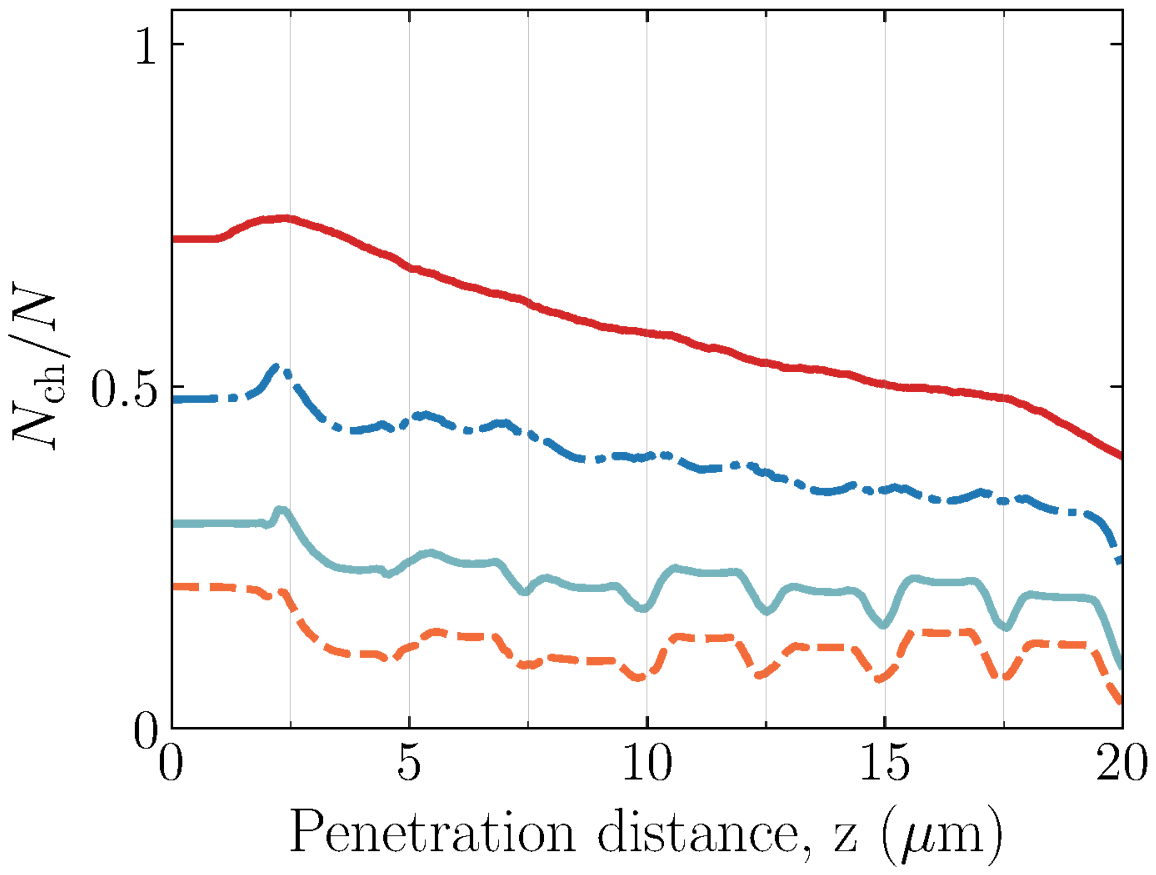}
\caption{Fractions of channeling electrons in straight and PB 
diamond (110) crystals.
Left: primary fractions of the accepted particles.
Right: fractions with account for the re-channeling.
Vertical lines, spaced by the half-period $\lamu/2$, mark 
the points of maximum curvature of the cosine bending
profile.}
\label{Figure04.fig}%{Figure04.fig}
\end{figure}
%%%%%%%%%%%%%%%%%%%%%

The interplay of several processes, occurring in the PBCs, lead to 
structural transformations in the part of the spectrum $\hbar\om \gtrsim 2$ MeV 
dominated by ChR.
This radiation is emitted by all particles experiencing channeling motion.
These include the accepted electrons as well as those 
re-channeled anywhere inside the crystal.
It was shown \cite{Korol_EtAl_EPJD_v71_174_2017} that periodicity 
in the bending enhances significantly the re-channeling rate
even in the limit of large bending curvatures $1/R_{\max}$. 
As shown, this is in contrast with the dechanneling rate which is virtually 
equal for both PB and uniformly bent crystalline structures. 
Left panel in Fig. \ref{Figure04.fig} shows the dependences of 
the fraction $N_{\text{ch}0}/N$ of the accepted electrons
which display the channeling motion at the distance $z$ from the entrance.
The fraction $N_{\text{ch}}/N$ of all electrons (the primarily and the 
re-channeled ones) that move in the channeling mode are shown on the right panel.
The vertical lines mark the points of maximum curvature of the bending profile.
At this distances, the effect of de-channeling is the largest leading to the minima of 
the channeling fraction $N_{\text{ch}}/N$. 
In contrast, at the distances where the curvature approaches the zero value, 
the re-channeling yields a significant increase in 
the number of channeling electrons. 
The dependence of $N_{\text{ch}}/N$ on $z$ allows one to calculate the 
total length $\Ltot$ of all channeling segments per a projectile.
The corresponding data are presented in the last column of 
Table \ref{Table_ep-data.C}.
%
% One can expect the intensity of ChR, $I_{\text{ChR}}^{(-)}$, to be proportional
% to $\Ltot$.

For the cosine profile of the bending, electrons enter the PBC in the point
of maximum curvature. 
The centrifugal force filters the particles with respect to the amplitudes $\ach$, so 
that similar to the positron case 
the \textit{accepted} particles oscillate with comparatively small amplitudes.
Anharmonicity of the electron channeling oscillations leads to a  monotonously decrease
of their frequency with $\ach$.
% (see, e.g., Refs. \cite{BakEtAl_NPB_v254_p491_1985,Korol_EtAl_NIMB_v387_p41_2016}). 
% Due to the anharmonic character of the electron channeling oscillations, their frequency 
% depends on the amplitude being a monotonously decreasing function of the latter 
% (see, e.g., Refs. \cite{BakEtAl_NPB_v254_p491_1985,Korol_EtAl_NIMB_v387_p41_2016}). 
Therefore, the low-frequency channeling oscillations are suppressed.
Since the emission frequency is related to $\Omch$ as $\om \approx 2\gamma^2\Omch$, 
one concludes that the ChR spectrum of the accepted particles in PBCs is 
blue-shifted in comparison with the straight one, and the shift increases 
with $a$. 
This feature is seen in Figs. \ref{Figure01.fig}(a)-(d) if one compares the positions
of the first maximum (marked with the upward arrow).
The account for the emission from the re-channeled electrons allows one to explain why
the ChR spectrum does not decrease with $a$ at so high rate as in the positron case.
Indeed, re-channeling events occur in the (nearly) straight parts of the PB channel
where the centrifugal force is small. 
Therefore, the amplitude $\ach$ of these particles is uniformly distributed 
within the interval $\ach \lesssim d/2$ giving rise to the emission into the 
whole interval of the ChR energy.
As a result, the low-energy part of the ChR is non-zero for all amplitudes considered,
and the ratio of the maximum values of the intensity with good accuracy
follows the ratio of the $\Ltot$ values.

Finally, let us comment on the additional structure (marked with downward arrows) 
which appears as a hump at ca 10 MeV in Fig. \ref{Figure01.fig}(b) and  
gradually shifts to higher energies as $a$ increases.
We attribute it to the coherent BrS radiation which appears in a PBC.
In a straight crystal, coherent BrS is emitted by projectiles which traverse the 
planes under the angle larger than Lindhard's critical angle 
% (the over-barrier particles) 
thus experiencing quasi-regular correlated collisions with the atoms
\cite{Sorensen_NIMB_v119_p2_1996}.
Its maximum is located well-above that of the ChR (in our case, it is beyond 30 MeV 
which is the largest abscissa in Fig. \ref{Figure01.fig}(a)).
Additional, less frequent, periodicity appears in the non-channeling segments 
of the simulated trajectories in the PBC due to the periodicity in bending. 
This periodic mode leads to the additional structure in the emitted spectra.

%%%%%%%%%%%%%%%%%%%%%%%%%%%%%%%%%%%%%%
In summary, 
the channeling and radiation of 855 MeV electrons and positrons in PB
diamond (110) crystalline structures has been simulated by means of accurate 
numerical procedure based on all-atom molecular dynamics.
A particular focus of the studies is on the impact of the bending amplitude $a$ on
the intensities of the emitted CUR and ChR.
We predict, in particular, that for both projectiles the intensity of CUR is a 
non-monotonous function of $a$, so that one can find the optimal periodic bending 
of the crystal planes which ensures the highest yield of CUR.
This result is important for the experimental studies of the radiation from CUs as
well as for designing and practical realization of PB 
crystalline structures as the key element of the novel, CU-based light sources.
Another important prediction concerns the possibility to manipulate with the 
ChR intensity by varying the positron beam energy $\E$ and the bending amplitude.
In particular, by a proper choice of $a$ one can decrease the ChR intensity by orders 
of magnitude but, at the same time, maintaining high yield of the CUR.
This effect can be utilize to reduce the radiative energy losses which in oriented
crystals are mainly due to ChR.
Low radiative energy losses maintain the stability of the CUR signal in thick crystals.

%%%%%%%%%%%%%%%%%%%%%%%%%%%%%%%%%%%%%%%%%%%%%%%%%
\vspace*{0.2cm}
The work was supported in part by the Alexander von Humboldt Foundation Linkage 
Grant and by the HORIZON 2020 RISE-PEARL project. 
We acknowledge the Supercomputing Center of 
Peter the Great Saint-Petersburg Polytechnic University 
(SPbPU) for providing the opportunities to 
carry out large-scale simulations.
We are grateful to Hartmut Backe and Werner Lauth (University of Mainz) for
useful discussions, to Alexey Verkhovtsev (DKFZ) for careful reading of the 
manuscript and useful comments,
and to Alexander Ustinov and Kiril Agapev (SPbPU) for the help in 
setting the simulations.

%%%%%%%%%%%%%%%%%%%%%%%%%%%%%%%
%\bibliography{bibliography_AK}
%%%%%%%%% 

\end{document}